\documentclass[12pt]{iopart}
\usepackage{psfrag}
\usepackage{graphicx}
\usepackage{iopams}

\newcommand{\be}{\begin{equation}}
\newcommand{\ee}{\end{equation}}
\newcommand{\beq}{\begin{eqnarray}}
\newcommand{\eeq}{\end{eqnarray}}

\def\lsim{\hbox{ \raise.35ex\rlap{$<$}\lower.6ex\hbox{$\sim$}\ }}
\def\gsim{\hbox{ \raise.35ex\rlap{$>$}\lower.6ex\hbox{$\sim$}\ }} 

\begin{document}
\title{Lattice Refining LQC from an Isotropic Embedding of Anisotropic
  Cosmology} 
\author{William Nelson\ead{nelson@gravity.psu.edu}}
\address{ Institute of Gravitation and the Cosmos, Penn State
University, State College, PA 16801, U.S.A.}
\author{Mairi Sakellariadou\ead{mairi.sakellariadou@kcl.ac.uk}}
\address{Department of
  Physics, King's College, University of London, Strand WC2R 2LS,
  London, U.K.}

\begin{abstract}
We demonstrate that it is possible to produce different isotropic
embeddings of anisotropic Loop Quantum Cosmology, resulting in
``lattice refinement'' of the isotropic system. To introduce the
general approach, we first use a simple model with only two anisotropic
directions. We then employ the specific case of a Bianchi~I model, to show
how the method extends to three-dimensional systems. To concisely
calculate the step-size of the resulting isotropic state, we define
the ``symmetric dual'' of states and operators, for the two- and
three-dimensional systems, respectively.

\end{abstract}

\pacs{04.60.Kz, 04.60.Pp, 98.80.Qc}

\maketitle

\section{Introduction}
Loop Quantum Gravity (LQG) quantises canonical General Relativity, in a
background independent and non-perturbative manner. It does so by
considering the three-dimensional spatial slice to be a network of
triads and then formulating the classical theory in terms of holonomies of
the triad connections, around closed edges of the network, and fluxes of
the triads, through the surfaces enclosed by these edges. Whilst the
full, inhomogeneous theory has yet to be fully developed, significant
progress has been made by applying the approach to symmetry reduced,
mini-superspace models. In particular, by applying these principles of
quantisation to Cosmology, one arrives at Loop Quantum Cosmology
(LQC), which has proved itself especially
successful~\cite{Bojowald:2003uh}. 

Within LQC it has been possible, on the one hand, to explicitly show
the avoidance of singularities that typically plague the classical
versions of these cosmological
theories~\cite{Singh:2009mz,Bojowald:2008ma} and, on the other hand,
to develop powerful effective theories, that capture first order
corrections to classical cosmology~\cite{lrr-2008-04}. In particular,
it has been shown that, in order for the cosmological theory to be
theoretically consistent~\cite{Ashtekar:2006wn,Corichi:2008zb}, the
physical volume (rather than the scale factor, or the triad component)
must be the fundamental variable used. This conclusion has been backed
up by various phenomenological~\cite{Nelson:2007wj,Nelson:2007um} and
theoretical~\cite{Nelson:2008vz} arguments.  However, it is important
to realise that since there is still lacking a complete derivation of
LQC from the fundamental inhomogeneous and anisotropic theory, it is
important to test LQC predictions for robustness
(see {e.g.}, Ref.~\cite{Corichi:2009pp}).

Various aspects of anisotropic cosmologies have been studied within
LQC in the past~\cite{Bojowald:2007ra,Maartens:2008dd,Chiou:2007sp},
however the first full and consistent quantisation of a Bianchi~I
cosmology (the simplest of anisotropic cosmological models) was
achieved in Ref.~\cite{Ashtekar:2009vc}~\footnote{Performing a von
  Neumann stability analysis of the Hamiltonian constraint equation
  valid for anisotropic Bianchi~I model, we have
  shown~\cite{Nelson:2009iv} that the the difference equation is
  unconditionally unstable.}.  Even more encouraging, the link back to
the underlying full theory (LQG) has been strengthened, by considering
the flux of the triads through surfaces consistent with the Bianchi~I
anisotropic case.  With the quantisation of the Bianchi~I model under
control, it is possible to ask whether LQC features, obtained within
the context of isotropic Friedmann-Lema\^{i}tre-Roberston-Walker
(FLRW) cosmology, are robust, at least with respect to this limited
extension of the symmetries of the system.

In this paper, we will show that one can choose among different
isotropic embeddings of the anisotropic Bianchi~I model. We will then
demonstrate that the choice of different embeddings has important
consequences for the precise form of discretisation, produced by the
loop quantisation procedure, in the isotropic sub-system.  Note that,
we use Bianchi~I as an example of an anisotropic model. In other
words, we do not imply that it is physically natural, nor that
our results are constrained to hold only for this type of an
anisotropic model.

In Section~2, the basics of LQC will be sketched, for
both isotropic and anisotropic models. In Section~\ref{sec:2d}, it
will be demonstrated that an isotropic embedding of an anisotropic
system can be realised in various ways, leading to consequences for
the form of the induced isotropic system. This will be first
demonstrated for a prototype two-dimensional system. In
Section~\ref{sec:sym_dual}, we will define a ``symmetric dual'' of
operators and states, to apply our formalism. In Section~5, the
approach will be extended to the three-dimensional Bianchi~I model and
it will be shown that, with certain additional caveats, the same
conclusions can be drawn.  Finally, we will demonstrate that ``lattice
refinement''~\cite{Nelson:2007um,Sakellariadou:2008qy} can be
motivated by alternative isotropic embeddings, while the difference
from the standard quantisation of isotropic cosmology can be viewed
simply as a different choice of basic observables.

\section{Basics formalism of LQG/LQC}

The full LQG theory is uniquely derived from the requirements that
the theory is diffeomorphism invariant and satisfies the ${\rm
  SU}(2)$ gauge freedom.  In both, the full theory and in LQC, the
fundamental variables are the holonomies of the ${\rm SU}(2)$
connection, $A^i_a$, along a given edge, $e$, and the corresponding
momentum, which turns out to be the flux of the triad through a
two-surface $S$.  The holonomies are given by
\be 
h_e(A) = {\cal P} \exp \int_e \dot{\gamma}^\mu(s) A^i_\mu
\left(\gamma(s) \right)\tau_i {\rm d}s~, 
\ee 
where ${\cal P}$ infers path ordering on the exponential,
$\gamma^\mu$ is the tangent vector along the edge $e$, and
$\tau_i$ are the basis of the ${\rm SU}(2)$ Lie algebra.
The corresponding momentum variable is
\be 
E(S,f) = \int_S\epsilon_{\rm
  abc} E^{{\rm c}i} f_i {\rm d}x^a {\rm d}x^b~, 
\ee 
where $f_i$ is an ${\rm SU}(2)$ valued test function and $E_i^{\rm a}$
is the densitised triad; $i,j,k, \cdots$ are ${\rm SU}(2)$
indices, whilst $a,b,c,\cdots$ are coordinate indices.

Restricting to isotropic and homogeneous cosmologies allows us to
consider only straight edges along integral curves of the basis
vectors, $X_i^a$, that produce the network.  Then the connection is
given by a (dynamic) multiple of the basis one forms, $\omega_a^i$,
namely $A^i_a = \tilde{c}(t) \omega^i_a$, whilst the triad is $E^a_i =
\sqrt{^0g} \tilde{p}(t) X^a_i$, where $^0g$ is the determinant of the
fiducial metric\footnote{The fiducial metric is a complication that
  arises only for an open universe and is used to define the volume to
  which spatial integrals are restricted to ensure they remain finite.
  Physical results should not depend on this
  volume~\cite{Vandersloot_PhD}.} and $\tilde{c}$ stands for the
connection component.  This is simply a consequence of the symmetries
imposed, since isotropy ensures that there can be no angular
dependence of the connection or triads, whilst homogeneity ensures
they are the same at every spatial point (but not necessarily on
different time slices).

With these symmetries, the holonomies become simply
\beq
\label{eq:hol} 
h_{\rm i}(A) &=& \exp \left[ \frac{-i\mu_0\sigma_{\rm i}}{2} \tilde{c}
  \right]~, \nonumber \\ 
&=& \cos \left( \frac{\mu_0 \tilde{c}}{2}
\right) +2\tau_i \sin \left( \frac{ \mu_0 \tilde{c}}{2} \right)~, 
\eeq
where $\tau_i$, the basis of the ${\rm SU}(2)$ Lie algebra, are related
to the Pauli matrices, $\sigma_i$, by $\tau_i=-i\sigma_i/2$ and
$\mu_0$ is the orientated length of the edge with respect to the
fiducial metric. It is this $\mu_0$ that is the ambiguity coming from
the fact that we have imposed homogeneity on our system. In a full
theory, the length of the edges would have some spectrum of values, in
which case $\mu_0$ would be the minimum such length, which has been
shown to be non-zero~\cite{Ashtekar:2003hd}.  Here we have treated
$\mu_0$ as a constant, however for consistent
quantisation~\cite{Corichi:2008zb,Nelson:2008vz,Nelson:2007wj}, it has
been shown that it must vary as $\mu_0 \sim \mu^{-1/2}$. In order to
include this varying parameter, one must change variables from the
triad component to the volume~\cite{Ashtekar:2006wn}, however here we will
briefly sketch the derivation for the constant $\mu_0$ case, with more
details given in Ref.~\cite{Ashtekar:2006wn}.

By analogue with the full theory, the kinematic Hilbert space is
extended via the Bohr compactification of the real
line~\cite{Ashtekar:2003hd}.  An orthonormal basis for this Hilbert
space is $\{ | \mu \rangle \}$, where
\be
 \langle c | \mu \rangle = e^{i\frac{\mu c}{2}}~.
\ee
Note that $c$ is a re-definition of the connection component, so that
the Poisson bracket of the connection component and the (re-defined)
triad component is independent of the volume of the fiducial cell,
namely $\{c,p\}=\kappa\gamma/3$, with $\kappa = 8 \pi G$ and $\gamma$
the Barbero-Immirzi parameter, fixed through a black hole entropy
calculation  and turns out to be $\gamma\approx 0.2375$. 

The triad operator acts on these basis states as 
\be
\label{eq:quan_p} 
\hat{p}|\mu \rangle=-i\frac{\kappa \gamma \hbar}{3} \widehat{
  \frac{\partial}{\partial c} }|\mu\rangle = \frac{\kappa \gamma
  \hbar}{6} |\mu| |\mu\rangle~.  
\ee 
Clearly, also the volume operator, $\hat{V}\equiv \hat{a}^3=
|\hat{p}|^{3/2}$, has eigenstates in this basis, namely
\be
\label{eq:vol}
\hat{V}|\mu\rangle = V_\mu|\mu\rangle =  \left(\frac{\kappa \gamma \hbar
 |\mu|}{ 6}\right)^{3/2} |\mu\rangle~. 
\ee
To calculate the eigenvalues of the inverse volume operator, the classical
expression~\cite{Ashtekar:2003hd},
\be\label{eq:inv}
p^{-1}= \left(p^{L-1}\right)^{1/\left(1-L\right)}= \left(\frac{3}{\kappa\gamma L}
 \{c,p^L\}\right)^{1/\left(1-L\right)}~,
\ee
is used. This is a classical identity independent of L, but will be
quantised to different operators for different L. Thus, L plays the
r\^ole of a quantisation ambiguity.

It is now possible to quantise the classical Hamiltonian constraint,
where the curvature is approximated via holonomies around a closed
curve (which can not be shrunk to zero, due to the ``area gap'', given
in terms of $\mu_0$; it occurs by analogue with LQG), which
gives
\be 
{\hat C}_{\rm grav} = \frac{2i}{ \kappa^2 \hbar \gamma^3 \mu_0^3} {\rm
  tr} \sum_{i,j,k} \epsilon^{ijk}\Bigl( \hat{h}_i^{(\mu_0)} \hat{h}_j^{(\mu_0)} 
\hat{h}_i^{(\mu_0)-1}
\hat{h}_j^{(\mu_0)-1} \Bigl[ \hat{h}_k^{(\mu_0)-1}, \hat{V} \Bigr] {\rm sgn}
\left(\hat{p}\right) \Bigr)~, 
\ee 
where, as in all quantum theories, there is an ambiguity in the factor
ordering~\cite{Nelson:2008vz}; other quantum ambiguities have been
fixed. Using Eq.~(\ref{eq:hol}), one can thus obtain a difference
equation~\cite{Vandersloot_PhD}, 
\beq\label{qee}
&&~~\Biggl[\Big| V_{\mu+5\mu_0}-V_{\mu+3\mu_0}\Big|+\Big|V_{\mu+\mu_0}
 - V_{\mu-\mu_0}\Big|\Biggr] \Psi_{\mu+4\mu_0}(\phi) -
 4\Big|V_{\mu+\mu_0}V_{\mu-\mu_0}\Big| \Psi_\mu(\phi) \nonumber \\
 &&+\Biggl[\Big|V_{\mu-3\mu_0}-
 V_{\mu-5\mu_0}\Big|+\Big|V_{\mu+\mu_0}- V_{\mu-\mu_0}\Big|\Biggr]
 \Psi_{\mu-4\mu_0}(\phi)\nonumber \\
 && =- \frac{4\kappa^2 \gamma^3 \hbar \mu_0^3}{3}
 {\mathcal H}_{\rm \phi}(\mu)\Psi_\mu(\phi)~,
\eeq
of step $4\mu_0$, for the wave-function coefficients
$\Psi\left(\mu\right)$, defined by
\be 
|\Psi\rangle = \sum_\mu \Psi\left(\mu\right) | \mu\rangle~.  
\ee
Note that, the matter Hamiltonian $\hat{\mathcal H}_{\rm \phi}$ is assumed
to act diagonally on the basis states with eigenvalue ${\mathcal
H}_{\rm \phi}(\mu)$.

By adapting the network to the symmetries of a Bianchi~I model, one
can similarly derive the quantum Hamiltonian for this anisotropic
system~\cite{Ashtekar:2009vc}. In this case, we have three triad
components $\left(p_1,p_2,p_3\right)$, along the three directions of
the anisotropic model.  The gravitational sector of the kinematic
Hilbert space consists of wave-functions $\Psi(p_1,p_2,p_3)$,
satisfying
\be
\Psi(p_1,p_2,p_3)=\Psi(|p_1|,|p_2|,|p_3|)~.
\ee
Neglecting the details of the derivation of the Hamiltonian
constraint, which are of no relevance for our present study, we draw
the attention of the reader to the fact that in both, the isotropic
and the Bianchi~I case, the holonomies act on a state as shift
vectors.

Let us define three dimensionless variables, $\lambda_i$, as 
\be
\lambda_i=\frac{{\rm
    sgn}(p_i)\sqrt{|p_i}|}{(4\pi|\gamma|\sqrt{\Delta}l_{\rm
    Pl}^3)^{1/3}} ~~\mbox{with}~~i=1,2,3~; 
\ee 
the quantum of area $\Delta l_{\rm Pl}^2$ denotes the physical
geometry, with $\Delta =4\pi\gamma\sqrt{3}$. The Hamiltonian
constraint for the wave-function $\Psi(\lambda_1,\lambda_2,\nu)$,
where $\nu=2\lambda_1\lambda_2\lambda_3$, is a difference equation. 
It reads~\cite{Ashtekar:2009vc}
\beq\label{diff-eq-aniso}
\partial_\phi^2\, \Psi(\lambda_1,\lambda_2,\nu;\phi) =& \frac{\pi G}
{2}\sqrt{\nu}\Big[(\nu+2)\sqrt{\nu+4}\,\Psi^+_4(\lambda_1,\lambda_2,\nu;\phi) 
\nonumber \\&
- (\nu+2)\sqrt
\nu\, \Psi^+_0( \lambda_1,\lambda_2,\nu;\phi)\nonumber \\& -(\nu-2)\sqrt \nu\,
\Psi^-_0(\lambda_1,\lambda_2,\nu;\phi)
\nonumber \\&
 + (\nu-2)
\sqrt{|\nu-4|}\,\Psi^-_4(\lambda_1,\lambda_2,\nu;\phi)\Big], 
\eeq
where $\Psi^\pm_{0,4}$ are defined as~\cite{Ashtekar:2009vc}
\beq
\Psi^\pm_0(\lambda_1,\lambda_2,\nu;\phi)=\nonumber\\
\ \ \ \ \Psi\left(\frac{\nu\pm2}{\nu}\cdot\lambda_1, \frac{\nu}{\nu\pm2}
\cdot\lambda_2,\nu;\phi\right)
+\Psi\left(\frac{\nu\pm2}{\nu}\cdot\lambda_1,\lambda_2,\nu;\phi \right)
\nonumber \\
\ \ +\Psi\left(\frac{\nu}{\nu\pm2}\cdot\lambda_1,\frac{\nu\pm2}{\nu}\cdot\lambda_2,
\nu;
\phi\right)+\Psi\left(\frac{\nu}{\nu\pm2}\cdot\lambda_1,\lambda_2,\nu;\phi\right)
\nonumber
\\
\ \ +
\Psi\left(\lambda_1,\frac{\nu}{\nu\pm2}\cdot\lambda_2,\nu;\phi\right)
+\Psi\left(\lambda_1,\frac{\nu\pm2}{\nu}\cdot\lambda_2,\nu;\phi\right)\, ,
\eeq
and
\beq
\Psi^\pm_4(\lambda_1,\lambda_2,\nu;\phi)= \nonumber\\
\ \ \ \ \Psi
\left(\frac{\nu\pm 4}{\nu\pm2}\cdot\lambda_1,\frac{\nu\pm2}{\nu}\cdot\lambda_2,
\nu\pm4;\phi\right)+\Psi\left(\frac{\nu\pm4}{\nu\pm2}\cdot\lambda_1,\lambda_2,
\nu\pm4;\phi\right)\nonumber\\
\ \ +\Psi\left(\frac{\nu\pm2}{\nu}\cdot\lambda_1,
\frac{\nu\pm4}{\nu\pm2}\cdot\lambda_2,\nu\pm4;\phi\right)+\Psi
\left(\frac{\nu\pm2}{\nu}\cdot\lambda_1, \lambda_2,\nu\pm4;\phi\right)\nonumber
\\
\ \ +\Psi\left(\lambda_1,\frac{\nu\pm2}{\nu}\cdot\lambda_2,
\nu\pm4;\phi\right)+\Psi\left(\lambda_1,\frac{\nu\pm4}{\nu\pm2}\cdot\lambda_2,
\nu\pm4;\phi\right),
\eeq
respectively. Note that, $\phi$ plays the r\^ole of {\sl internal}
time, exactly as in the isotropic case.  Thus, as one can easily check
from Eq.~(\ref{diff-eq-aniso}), as far as the $\nu$ dependence is
concerned, the steps are uniform, since the
argument of the wave-function involves $\nu-4, \nu, \nu+4$, exactly as
in the isotropic case.  However, the dependence on $\lambda_1,
\lambda_2$ is quite complicated, technically, to deal with.

Writing the step-size in terms of the triad components, implies
\beq\label{eq:step-size}
\Psi_{\rm iso}\left(p\right) \rightarrow  \Psi_{\rm iso}\left( p \pm
\frac{K}{\sqrt{|p|}} \right)\nonumber
\\ 
\mbox{and}\ \ \ \ \Psi\left(p_1,p_2,p_3\right) \rightarrow
\left\{ \begin{array}{c}
\Psi \left( p_1 \pm
\tilde{K}\sqrt{|\frac{p_1}{p_2 p_3}|},p_2,p_3 \right) \\
\Psi \left( p_1,p_2\pm
\tilde{K}\sqrt{|\frac{p_2}{p_1 p_3}|},p_3 \right) \\
\ \ \Psi \left( p_1,p_2,p_3\pm
\tilde{K}\sqrt{|\frac{p_3}{p_1 p_2}|}\right)~,
\end{array} \right.
\eeq
for some suitable constants $K$ and $\tilde{K}$ that are proportional
to $\sqrt{\Delta} l_{\rm Pl}^3$~\cite{Ashtekar:2009vc}. Note that,
typically the $\tilde K$ constants, appearing in the argument of the
anisotropic wave-function, will be different.

In both, the isotropic and anisotropic cases, ``consistency'' is derived
entirely within the model, {\sl i.e.}, the symmetries of the model are
fixed from the outset and ``consistency'' is looked for, under these
assumptions. This, of course, does not ensure that the model remains
``consistent'', if it derived from a less symmetric approximation.  In
particular, the isotropic case is consistently quantised for the
step-size of the holonomies given in Eq.~(\ref{eq:step-size}), only if
one does not allow additional degree of freedom (and similarly for the
Bianchi~I model), however this does not ensure that an isotropic
embedding of the Bianchi~I model will have the same step-size as the
isotropic case, given in Eq.~(\ref{eq:step-size}). However, with both
models fully quantised, it is possible to consider such an isotropic
embedding of the anisotropic Bianchi~I model and explicitly evaluate
the type of step-size that is induced on the effective isotropic
system.

This will be done in some detail below, but one can
immediately see that setting $p_1 = p_2 = p_3=p$, will produce the
expected form of step-size, however this embedding procedure does not
lead to a viable isotropic theory, instead one has to use a projection 
of the anisotropic degrees of freedom~\cite{Ashtekar:2009vc}.
The reason for this is the following: for $p_1 = p_2 = p_3=p$, we have
$\lambda_1 = \lambda_2 = \lambda_3 = (\nu/2)^{1/3}$.
Thus, the state which satisfies $p_1 = p_2 = p_3=p$ is
$$ \Psi\left( \left(\frac{\nu}{2}\right)^{1/3},
\left(\frac{\nu}{2}\right)^{1/3}, \nu\right)~.$$ From
Eq.~(\ref{diff-eq-aniso}), we see that applying the Hamiltonian to
such a state involves states of the form
$\tilde{\Psi}\left(\tilde{\lambda}_1,\tilde{\lambda}_2,\tilde{\nu}\right)~,$
which do not satisfy $\tilde{\lambda}_1 = \tilde{\lambda}_2 =
\tilde{\lambda}_3 = (\tilde{\nu}/2)^{1/3}$.  Thus, states for which
$p_1 = p_2 = p_3=p$, do not form a {\sl super-selection} of states ({\sl
  i.e.}, the Hamiltonian does not preserve this property of states).

The purpose of this paper is to show that it is, in fact, possible to
find a {\sl super-selection} of states that are all (measured to be) 
isotropic. In particular, we will show that there are embeddings for
which the expectation values of the variables $p_1$, $p_2$ and $p_3$
are equal, and that are preserved under the action of the
Hamiltonian. The Hamiltonian applied to such states does not produce
the standard FLRW Hamiltonian of LQC, rather one finds a Hamiltonian
constraint in which the discreteness length varies. Thus, one may view
this as a form of lattice refinement on the isotropic sub-space. It is
important to note however that the typical forms of lattice refining
Hamiltonian used in the literature are not the same as those produced
here; there is instead  a simplified attempt to capture the missing
underlying degrees of freedom (in this case the underlying anisotropic
degrees of freedom).

\section{The two-dimensional case}\label{sec:2d}

To demonstrate the approach, let us first consider an anisotropic
wave-function that depends only on two directions, labelled $1,2$. The
basic operators that are quantised are $\hat{p}_i$ and $\hat{h}_i$,
with $i=1,2$. Note that $p_i = a_i^2$, while $h_i$ denote the
holonomies.

In general, the wave-function is anisotropic, namely
\be
 |\Psi\left(p_1,p_2\right)\rangle \neq |\Psi \left( p_2,p_1\right) \rangle~.
\ee
To be clear, the action of the operators $\hat{p}_i$ is
\beq
\hat{p}_1 |\Psi\left( p_1,p_2\right) \rangle &=& p_1
|\Psi\left(p_1,p_2\right) \rangle\nonumber\\ \hat{p}_2 |\Psi\left(
p_1,p_2\right) \rangle &=& p_2 |\Psi\left(p_1,p_2\right)
\rangle
\eeq
and
\beq
\hat{p}_1 |\Psi\left( p_2,p_1\right) \rangle &=&
p_2 |\Psi\left(p_2,p_1\right) \rangle\nonumber\\ \hat{p}_2
|\Psi\left( p_2,p_1\right) \rangle &=& p_1 |\Psi\left(p_2,p_1\right)
\rangle~,\nonumber\\
\eeq
and the action of the holonomies $\hat{h}_i$ on a states $|\Psi\left(p_1,p_2\right)\rangle$
and  $|\Psi\left(p_2,p_1\right)\rangle$ are,
\beq 
\label{eq:3} 
\ \ \ \ \ \ \ \ \ \hat{h}_1 |\Psi\left( p_1,p_2\right)\rangle &=&
|\Psi\left( p_1 + \delta_1\left(p_1,p_2\right), p_2\right)\rangle
\nonumber \\
 \mbox{and}\ \ \ \ \hat{h}_2 |\Psi\left( p_1,p_2\right)\rangle &=&
|\Psi\left( p_1, p_2+ \delta_2\left(p_1,p_2\right) \right)\rangle~,
\nonumber \\ 
\eeq 
and
\beq
\ \ \ \ \ \ \ \ \ \hat{h}_1 |\Psi\left( p_2,p_1\right)\rangle &=&
|\Psi\left( p_2 + \delta_1\left(p_1,p_2\right), p_1\right)\rangle
\nonumber \\
 \mbox{and}\ \ \ \ \hat{h}_2 |\Psi\left( p_2,p_1\right)\rangle &=&
|\Psi\left( p_2, p_1+ \delta_2\left(p_1,p_2\right) \right)\rangle~,
\nonumber \\ 
\eeq
where $\delta_i$ is potentially a function of both $p_1$ and $p_2$.

\subsection{Standard isotropic embedding}

Typically, one can go to the isotropic case by defining the operator
$\hat{Q}_{\rm sym}$, as
\be 
\hat{Q}_{\rm sym} |\Psi\left(p_1,p_2\right)\rangle \equiv
|\Psi\left( p_1, p_1\right)\rangle = |\Psi\left(p_2,p_2\right)\rangle
= |\Psi\left(p,p\right)\rangle~.
\ee
In other words, the operator $\hat{Q}_{\rm sym}$ sets $p_1 = p_2$. It
is easy to see that operating singly with either $\hat{h}_1$, or
$\hat{h}_2$, on $|\Psi\left( p_1,p_2\right)\rangle$, before operating
with $\hat{Q}_{\rm sym}$, implies $\delta_i = 0$, {\sl e.g.},
\beq
 \hat{Q}_{\rm sym} \left[ \hat{h}_1 |\Psi\left( p_1,p_2\right)\rangle
   \right] &=& \hat{Q}_{\rm iso} |\Psi\left( p_1+\delta_1,p_2\right)
 \rangle \nonumber \\ &=& |\Psi\left( p_1+\delta_1,p_1+\delta_1\right)
 \rangle \nonumber \\ &=& |\Psi \left( p_2,p_2\right) \rangle\nonumber
 \\ \ \ \ \  \ \ \ \ \ \ \ \ \ \Rightarrow  \ \ \ \ p_1 + \delta_1 &=& p_2~,
\eeq
at least for singly defined wave-functions.  Thus, the operation is not
conserved under the action of a single holonomy, unless $\delta_1 =0$.
This can be understood easily, since a general anisotropic state can
be always made isotropic by setting $p_1 = p_2$, but if we change
either $p_1$ or $p_2$, then the condition required to now make them
equal will, of course, change.

If we consider only isotropic applications of the holonomies, {\sl
  e.g.}, $\hat{h}_1\hat{h}_2$, then the operation can be conserved
under the action of $\hat{Q}_{\rm sym}$, provided $\delta_1 =
\delta_2$. To be explicit, 
\beq \hat{Q}_{\rm sym} \left[ \hat{h}_1 \hat{h}_2
  |\Psi\left(p_1,p_2\right)\rangle \right] =\hat{Q}_{\rm sym} |\Psi
\left( p_1 + \delta_1, p_2+\delta_2 \right)\rangle \nonumber
\\ \ \ \ \ \ \ \ \ \ \ \ \ \ \ \ \ \ \ \ \ \ \ \ \ \ \ \ \ \ =
|\Psi\left(p_1+\delta_1,p_1+\delta_1\right)\rangle
\nonumber\\ \ \ \ \ \ \ \ \ \ \ \ \ \ \ \ \ \ \ \ \ \ \ \ \ \ \ \ \ \ =
|\Psi \left( p_2+\delta_2,p_2+\delta_2\right)\rangle~, \nonumber
\\ \ \ \ \ \ \ \ \ \ \ \ \ \ \Rightarrow \ \ \ \ p_1 +\delta_1 = p_2
+ \delta_2~, \eeq which preserves the condition $p_1 = p_2$ iff
$\delta_1 = \delta_2$, for singly defined wave-functions.

So, we found that the value of $p$ in either direction is the same and
it changes in the same way, namely it is isotropic. Notice that, for
consistency, this requires that we only consider isotropic
applications of the holonomy operators ({\sl e.g.}, the Hamiltonian),
which is physically acceptable since if our system is to end up
isotropic, then it must evolve isotropically. However, in the Bianchi
Hamiltonian (Eq.~(\ref{diff-eq-aniso})) the holonomies act via
multiplication, rather than addition, of the variables thus,
\beq\label{eq:1_new} \hat{Q}_{\rm sym} \left[ \hat{h}_1 \hat{h}_2
  |\Psi\left(p_1,p_2\right)\rangle \right] = \hat{Q}_{\rm sym} |\Psi
\left( \delta_1 p_1,\delta_2 p_2 \right)\rangle \nonumber
\\ \ \ \ \ \ \ \ \ \ \ \ \ \ \ \ \ \ \ \ \ \ \ \ \ \ \ \ \ \ =
|\Psi\left(\delta_1p_1,\delta_1p_1\right)\rangle
\nonumber\\ \ \ \ \ \ \ \ \ \ \ \ \ \ \ \ \ \ \ \ \ \ \ \ \ \ \ \ \ \ =
|\Psi \left( \delta_2 p_2,\delta_2 p_2\right)\rangle~, \nonumber
\\ \ \ \ \ \ \ \ \ \ \ \ \ \ \Rightarrow \ \ \ \ \delta_1 p_1 =
\delta_2 p_2~, \eeq
which again preserves the condition $p_1 = p_2$ iff $\delta_1 = \delta_2$,
(again for singly defined wave-functions). The full Bianchi Hamiltonian
(Eq.~(\ref{diff-eq-aniso})) contains linear combinations of operations of
the form of Eq.~(\ref{eq:1_new}) that do not satisfy $\delta_1 = \delta_2$.
Thus, the condition $p_1= p_2$ is not preserved. In order for the standard
``improved dynamics'' of the isotropic model to be recovered,
 one must project out the anisotropic degrees of freedom~\cite{Ashtekar:2009vc}.

\subsection{Alternative isotropic embedding}
\label{subsec:alternative}

There is also another approach one could follow to forming an
isotropic system from the fully (two-dimensional) anisotropic state
$|\Psi\left(p_1,p_2\right)\rangle$. Let us define the symmetrisation
operator by
\be
 \hat{Q}_{\rm sym} |\Psi\left(p_1,p_2\right) \rangle = \frac{1}{2}
 \Bigl[ |\Psi\left(p_1, p_2\right)\rangle + |\Psi\left(p_2,p_1\right)
   \rangle \Bigr]~.
\ee
Thus, whilst (for example) $\hat{p}_1 \left[ \hat{Q}_{\rm sym}
  |\Psi\left(p_1,p_2\right)\rangle \right]$ does not appear to be
isotropic, the physically expectation value of $\hat{p}_1$ between two
such states is. To be clear,
\be
 \hat{p}_1 \Bigl[ \hat{Q}_{\rm sym} |\Psi\left(p_1,p_2\right)\rangle
   \Bigr] = \frac{1}{2}\Bigl[ p_1|\Psi\left(p_1,p_2\right) \rangle +
   p_2 |\Psi\left(p_2,p_1\right) \rangle \Bigr]~,
\ee
is not symmetric, however the inner product,
\beq
&& \langle \Psi \left(p_1,p_2\right) \hat{Q}_{\rm sym} | \hat{p}_1 |
\hat{Q}_{\rm sym} \Psi\left(p_1,p_2\right) \rangle  \nonumber
\\ &=&\frac{1}{4}\Bigl[ p_1 + p_1 \langle \Psi \left( p_1,p_2\right)
  |\Psi\left(p_2,p_1\right) \rangle + p_2 \langle \Psi\left(
  p_2,p_1\right)|\Psi\left(p_1,p_2\right)\rangle + p_2 \Bigr]
\nonumber \\ &=& \frac{\left(p_1 + p_2\right)}{4}\left[ 1 + \langle
  \Psi \left( p_1,p_2\right) | \Psi\left(p_2,p_1\right) \rangle
  \right] \nonumber \\ &=& \langle \Psi \left(p_1,p_2\right)
\hat{Q}_{\rm sym} | \hat{p}_2 | \hat{Q}_{\rm sym}
\Psi\left(p_1,p_2\right) \rangle~,
\eeq
is clearly symmetric. Note that, we used $\langle X|Y\rangle=\langle 
Y|X\rangle$ and the notation
\be
 \langle \Psi \left(p_1,p_2\right) \hat{Q}_{\rm sym} | \equiv \left(
 \hat{Q}_{\rm sym} | \Psi\left( p_1,p_2\right)\rangle
 \right)^{\dagger}~.
\ee
Thus, the measured scale factor along either direction is the same. 

We can now consider what the action of an holonomy is and, in
particular, whether it is conserved under the symmetrisation operator.
Consider, for example, the operation of $\hat{h}_1$:
\be 
\hat{Q}_{\rm sym} \left[ \hat{h}_1|\Psi\left( p_1,p_2\right) \rangle
  \right] = \frac{1}{2} \Bigl[ |\Psi\left(p_1+\delta_1, p_2 \right)
  \rangle + |\Psi\left(p_2,p_1+\delta_1\right) \rangle \Bigr]~.  
\ee
We can again check that this state is symmetric, in that the
expectation values of the two $p_i$'s are the same, namely 
\beq 
&& \langle
\Psi\left(p_1,p_2\right) \hat{h}_1 \hat{Q}_{\rm sym} | \hat{p}_1 |
\hat{Q}_{\rm sym} \hat{h}_1 \Psi\left(p_1,p_2\right)\rangle 
\nonumber \\
&=&\frac{p_1 + p_2 + \delta_1}{4}\Bigl[ 1 + \langle \Psi\left(
  p_1+\delta_1,p_2\right)| \Psi\left(p_2,p_1+\delta_1\right)\rangle
  \Bigr]~, \nonumber \\
&=& \langle \Psi\left(p_1,p_2\right) \hat{h}_1 \hat{Q}_{\rm sym} | \hat{p}_2 |
\hat{Q}_{\rm sym} \hat{h}_1 \Psi\left(p_1,p_2\right)\rangle~.
\eeq

The above conclusion is however only valid, if the symmetrisation
operator is always the last one to be taken on the state. This has
some undesirable features, in particular,
\be
 \hat{O} \left(\hat{Q}_{\rm sym} | \Psi \rangle \right) \neq
\hat{Q}_{\rm sym} \left( \hat{O} |\Psi\rangle \right)~,
\ee
namely that the symmetrisation operation does not commute with other
operators, which is problematic, since the operation of a generic
operator ($\hat{O}$) on a symmetric state, does not, on its own, lead
to a symmetric state. In the next section, we will show how that this
problem is avoided provided we are interested only in ``symmetric operators'' .
In order to define such operators, it is useful to define a
process we refer to as the ``symmetric dual''.

\section{The ``symmetric dual''}\label{sec:sym_dual}

Here, we will define the ``symmetric dual'' of operators and states.
Note that, the symmetrised version of an operator must be used even if
$p_1 = p_2$ (we refer the reader to the previous section), since one
has to ensure that only symmetric products of the holonomies are
taken.

Define the symmetrised version of a state, $|\Psi\rangle_{\rm sym}$, as
\beq
 |\Psi\rangle_{\rm sym} &\equiv& \frac{1}{A} \bigl[ |\Psi\rangle +
   |\Psi \rangle^{\rm S} \bigr]~,\\
 &=& \frac{1}{A} \bigl[ |\Psi\left(p_1,p_2\right) \rangle + | \Psi
   \left( p_2,p_1\right)\rangle \bigr]~,
\eeq
where by $|\Psi\rangle^{\rm S}$ we denote the symmetric ``dual'' of a state
$|\Psi\left(p_1,p_2\right)\rangle$. Similarly, define the symmetrised
version of an operator, $\hat{O}_{\rm sym}$, as 
\be\label{eq:sym_op}
\hat{O}_{\rm sym} \equiv \frac{1}{B} \left[ \hat{O} + \hat{O}^{\rm S} \right]~,
\ee 
where the operation of the ``dual'' operator is defined by
requiring, 
\be 
\left( \hat{O} |\Psi\rangle \right)^{\rm S} = \hat{O}^{\rm S}
|\Psi\rangle^{\rm S}~, 
\ee 
and noting that $\left( |\Psi \rangle^{\rm S} \right)^{\rm S}
= |\Psi\rangle$, by definition. 
We thus get 
\beq 
\hat{O}_{\rm sym}|\Psi\rangle_{\rm sym} &=& \frac{1}{AB} \left[
  \hat{O}|\Psi\rangle + \hat{O}|\Psi \rangle^{\rm S} + \hat{O}^{\rm S}|\Psi\rangle
  + \hat{O}^{\rm S}|\Psi\rangle^{\rm S} \right]~, \\
&=& \frac{1}{A} \left[ | \tilde{\Psi} \rangle + |\tilde{\Psi}\rangle^{\rm S}
  \right]~, 
\eeq 
where 
\be |\tilde{\Psi}\rangle = \frac{1}{B} \left[
  \hat{O}|\Psi\rangle + \hat{O}^{\rm S} |\Psi\rangle \right]~.  
\ee
\begin{figure}
 \begin{center}
  \includegraphics[scale=0.75]{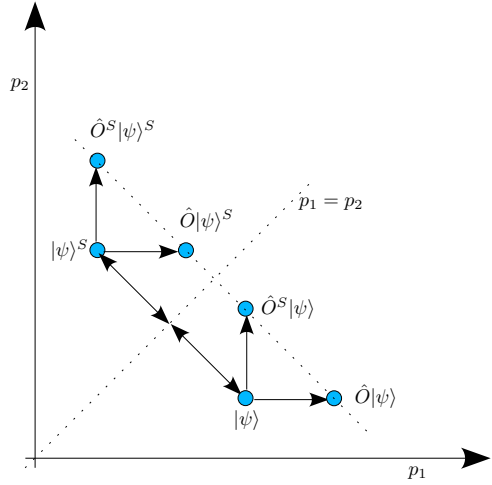}
  \caption{\label{fig:1} Visualisation of the symmetrisation of states.}
 \end{center}
\end{figure}
Thus, we find that the operation of a symmetric operator on a symmetric
state is (as expected) a symmetric state. This can be visualised on
the two-dimensional $(p_1,p_2)$ plane. The case of $p_1=p_2$ is the
diagonal line, and if we restrict ourselves to states that lie on this
line, then we must also restrict to operators that act {\sl
  diagonally}, as one can see in Fig.~\ref{fig:1}. However, if we
consider a symmetric superposition of states, then in order for the
operation of a general operator $\hat{O}$ to give another symmetric
state, one should require the operation to be symmetric. One easy way
of doing this, is to define the dual operator, $\hat{O}^{\rm S}$, and act
with the symmetric combination $\hat{O}_{\rm sym} \sim
(\hat{O}+\hat{O}^{\rm S})$. Note that, this procedure is by no
means unique, however such symmetric operators maintain the symmetric subspace
made up of states $|\Psi\rangle_{\rm sym}$, {\sl i.e.}, if we restrict ourselves
to considering only such operators, then we have a {\it super-selection}
of states $|\Psi \rangle_{\rm sym}$.

In  Fig.~\ref{fig:1}, we have taken the operator $\hat{O}$ to act on a
state by shifting $\left(p_1,p_2\right)$. This is, of course,
motivated by the fact that we want to consider holonomies, which we
will assume act as shift operators, as in Eq.~(\ref{eq:3}).  Consider
the restrictions placed on $\delta_1\left(p_1,p_2\right)$ and
$\delta_2\left( p_1,p_2\right)$, by requiring that
\be
 \hat{h}_1\hat{h}_2 = \hat{O} = \hat{O}^{\rm S}~,
\ee
{\sl i.e.}, that the action of a holonomy along each direction, is a
symmetric operation in the sense defined above. As we will see, does
not mean that the state is ``shifted'' diagonally. We have 
$$
\hat{h}_1\hat{h}_2 |\Psi\rangle = \hat{h}_1\hat{h}_2
|\Psi\left(p_1,p_2\right)\rangle = |\Psi\left(
p_1+\delta_1\left(p_1,p_2\right),p_2+\delta_2\left(p_1,p_2\right)
\right) \rangle \nonumber \\
$$
and
$$
\hat{h}_1\hat{h}_2 |\Psi\rangle^S = \hat{h}_1\hat{h}_2
 |\Psi\left(p_2,p_1\right)\rangle = |\Psi\left(
 p_2+\delta_1\left(p_2,p_1\right),p_1+\delta_2\left(p_2,p_1\right)
 \right) \rangle~. \nonumber \\
$$
Thus, to satisfy
\be \left[ \hat{O} |\Psi\rangle\right]^{\rm S} = \hat{O}^{\rm S}
|\Psi\rangle^{\rm S} = \hat{O}|\Psi\rangle^{\rm S}~, \ee
one should require that
\beq
 |\Psi\left( p_2 + \delta_2\left(p_1,p_2\right),
 p_1+\delta_1\left(p_1,p_2\right) \right)\rangle =\nonumber\\ 
\ \ \ \ \ \ \ \ \ \ \ \ \ \ \ \ \ \ \ \ \ \ \ \ |\Psi\left( p_2 +
 \delta_1\left(p_2,p_1\right), p_1+\delta_2\left(p_2,p_1\right)
 \right)\rangle~,
\eeq
which implies
\be\label{eq:1}
 \delta_2\left(p_1,p_2\right) = \delta_1\left(p_2,p_1\right)~.
\ee 
For illustration, let us consider some simple examples. For $\delta_1
= \delta_2 = {\rm const.}$, the pair of holonomies acts diagonally,
with a constant step.  A more interesting example is 
$$\delta_1 = \frac{1}{\sqrt{p_1}} = \frac{1}{\sqrt{p_2}} = \delta_2~,$$ 
which is valid only when $p_1=p_2$, namely on the diagonal line. This
is the analogue of the ``new'' quantisation (or ``improved''
quantisation) approach. Motivated by the three-dimensional anisotropic
results of Ref.~\cite{Ashtekar:2009vc}, one can take
$$
\delta_1 \left(p_1,p_2 \right) =
\sqrt{\frac{p_1}{p_2}} \ \ \ , \ \ \ \delta_2 \left(p_1,p_2 \right) =
\sqrt{\frac{p_2}{p_1}}~.
$$ 
Clearly there are many other cases one may consider.

The question then is, what would be the measured step-size between
states separated by this symmetric operator.
We have
\beq _{\rm sym}\langle
\Psi\left(p_1,p_2\right)|\hat{p_1}|\Psi\left(p_1,p_2\right)\rangle_{\rm
  sym} &=& _{\rm sym}\langle
\Psi\left(p_1,p_2\right)|\hat{p_2}|\Psi\left(p_1,p_2\right)\rangle_{\rm
  sym} \nonumber \\ &=&\frac{p_1+p_2}{A^2}\Bigl[ 1 + \langle
  \Psi\left( p_1,p_2\right)|\Psi \left( p_2,p_1\right)
  \rangle \Bigr] \nonumber \\ &=&\frac{p_1+p_2}{A^2} \Bigl[ 1 +
  \langle \Psi|\Psi\rangle^{\rm S} \Bigr]
\nonumber\\ &=& p_1 + p_2~, \eeq
where in the last line, we used the fact that the symmetrised state should be
normalised, and we have chosen
\be\label{normalisation}
\langle \Psi| \Psi\rangle^{\rm S} =
\langle \Psi\left(p_1,p_2\right)|\Psi\left(p_2,p_1\right)\rangle 
= A^2 - 1~.
\ee
Defining,
\be
|\psi\left(p_1,p_2\right)\rangle_{\rm sym}\equiv\hat{h}_1\hat{h}_2|\Psi 
\left(p_1,p_2\right) \rangle_{\rm sym}~,
\ee
and taking the inner product, we find
\beq _{\rm sym}\langle \psi\left(p_1,p_2\right)| \hat{p}_1
|\psi\left(p_1,p_2\right)\rangle_{\rm sym} = _{\rm sym}\langle
\psi\left(p_1,p_2\right)| \hat{p}_2
|\psi\left(p_1,p_2\right)\rangle_{\rm sym} \nonumber
\\ =\frac{p_1+p_2+\delta_1\left(p_1,p_2\right)
  +\delta_1\left(p_2,p_1\right)}{A^2}  \left[ 1 +
  \Big\langle \Psi\left( \tilde{p_1},\tilde{p_2}\right)
  \Big|\Psi \left( \tilde{p_2},\tilde{p_1}\right)
  \Big\rangle \right], \eeq
where $$\left( \tilde{p_1},\tilde{p_2}\right) = \left(p_1 +
\delta_1\left(p_1,p_2\right),p_2+\delta_2\left(p_1,p_2\right) \right)$$
and
$$\left( \tilde{p_2},\tilde{p_1}\right) = \left(p_2 +
\delta_1\left(p_2,p_1\right),p_1+\delta_2\left(p_2,p_1\right) \right).$$
We can then evaluate what the measured step-size, $\Delta_{\rm sym}$, would be:
\beq\label{eq:Delta}
\Delta_{\rm sym} &\equiv& _{\rm sym} \langle
\psi\left(p_1,p_2\right)|
\hat{p}_i |\psi\left(p_1,p_2\right)\rangle_{\rm
  sym}
- _{\rm sym}\langle \Psi\left(p_1,p_2\right) | \hat{p}_i | \Psi
\left(p_1,p_2\right) \rangle_{\rm sym} \nonumber \\
&=&
\left(p_1 + p_2\right) \left(X-1\right) + \left(\delta_1 \left(p_1,p_2\right) +
\delta_2\left(p_1,p_2\right)\right)X~,
\eeq
where $X$ is defined as
\be X = \frac{1 + \langle
  \tilde{\Psi}\left(\tilde{p}_1,\tilde{p}_2\right) |
  \tilde{\Psi}\left(\tilde{p}_2,\tilde{p}_1\right)\rangle}{1 + \langle
  \tilde{\Psi} \left( p_1,p_2\right) |
  \tilde{\Psi}\left(p_2,p_1\right)\rangle }~.  \ee
Thus, we have
\be X= \frac{1 + \langle \tilde{\chi} | \tilde{\chi} \rangle^{\rm
    S}}{1 + \langle \tilde{\Psi} | \tilde{\Psi}\rangle^{\rm S} }~, \ee
and we used $|\tilde{\chi}\rangle \equiv
|\tilde{\Psi}\left(\tilde{p_1},\tilde{p_2} \right)\rangle$ to
emphasise the possibility of a difference from
$|\tilde{\Psi}\left(p_1,p_2\right)\rangle$.  The normalisation of the
symmetrised state gives the denominator (see,
Eq.~(\ref{normalisation})), but we get $X=1$ iff the norm of every
state (or every state in the super-selection, if one is possible) is
the same.  In particular, if we were to restrict ourselves to
$p_1=p_2\equiv p$, this would imply that $\tilde{p}_1 = \tilde{p}_2$
and hence that $|\tilde{\chi} \rangle = | \tilde{\chi} \rangle^{\rm S}$ and
$|\tilde{\Psi}\rangle = |\tilde{\Psi}\rangle^{\rm S}$.  This gives $X=1$,
which means $\Delta = \delta_1\left(p_1,p_2\right) +
\delta_1\left(p_2,p_1\right)= 2\delta\left(p\right)$.  However, for
general (``symmetric dual'') states, $X$ can be greater, or less, than
unity. Thus, the step-size would include a different dependence on
$p_1$ and $p_2$; it can be seen as a ``lattice refinement'' analogue.

In Section~\ref{sec:3D}, we will extend to three directions and
explicitly use the Bianchi~I model of Ref.~\cite{Ashtekar:2009vc} to
evaluate the step-sizes for the two different symmetrisation schemes.
Indeed, the above can be used directly for the case that two of the
directions are set equal to each other. For example, if the
restriction $p_1=p_2 \neq p_3$ is made, then we are restricted to a
sub-class of Bianchi~I models, which are essentially two-dimensional.
Such cylindrically symmetric models would be useful for investigating
black holes, where the expectation value of $\hat{p}$ is different
along the radial and angular directions.

Intuitively, what we have is that the use of superpositions to
symmetrise a state, means that (for symmetric operators, such as pairs
of holonomies with the correct type of step) we can find symmetric
states which however have different step-sizes than in the standard case.
Essentially, the standard case is just a special case of the above
procedure, when the two states used in the superposition are the same.

To be clear, we have chosen to normalise the states, such that
\be
\langle \Psi \left( p_1,p_2\right) |\Psi\left(p_2,p_1\right)\rangle =
A^2 - 1~.  
\ee 
However, the restriction $\delta_1\left(p_1,p_2\right) =
\delta_2\left(p_2,p_1\right)$, is not sufficient to simultaneously specify 
the normalisation of
$$
\langle \Psi\left[ p_1 +
   \delta_1\left(p_1,p_2\right),p_2+\delta_2\left(p_1,p_2\right)
   \right]| \Psi \left[ p_2 + \delta_1\left( p_2,p_1\right), p_1+
   \delta_2\left(p_2,p_1\right)\right] \rangle~.  
$$
For $p_1=p_2$,
\beq
\Delta_{\rm sym} &=& \delta_1\left(p_1,p_2\right) + 
\delta_2\left(p_1,p_2\right)\nonumber\\
&=& 2\delta\left(p\right)~,
\eeq
where the last equality follows from the fact that
$\delta_1\left(p_1,p_2\right) = \delta_2\left(p_2,p_1\right)
\equiv\delta\left(p\right)$. For any other choice of
$\left(p_1,p_2\right)$, such that $p_1 \neq p_2$, this is not the
case. Whilst we cannot rule out the possibility that the (symmetric)
state evolves towards a state made up of a superposition of two even 
more anisotropic states, this seems unlikely.  

If we take that, at least for large $p_1+p_2$, we have $|\Psi
\left(p_1,p_2\right) \rangle \rightarrow |\Psi
\left(p,p\right)\rangle$, then we expect $X\searrow 1$ ({\it i.e.},
the inner product of the ``symmetric dual'' states tends to $1$ from
above; it can be understood by thinking of the states as vectors in a
vector space). Thus,
\be 
\Delta_{\rm sym} \geq \delta_1\left(p_1,p_2\right) +
\delta_2\left(p_1,p_2\right)~, 
\ee
whilst tending to equality as $p_1$ and $p_2$ become large and tend to
each other. 

For the case of an isotropic system, made up of ever more anisotropic
states, the converse is true, namely,
\be 
\Delta_{\rm sym} \leq \delta_1\left(p_1,p_2\right) +
\delta_2\left(p_1,p_2\right)~, 
\ee 
whilst not necessarily tending to equality as $p_1$ and $p_2$ become
large.  Note that here we also require Eq.~(\ref{eq:1}) to hold, which
is what ensures that the measured step-size, $\Delta_{\rm sym}$, is
indeed symmetric under interchange of $p_1, p_2$. 

In conclusion, in general the step-size can be either larger or
smaller (and has a different dependence on $\left(p_1,p_2\right)$)
than in the $p_1 = p_2$ case. For the case of a state made up of a
superposition of ever more isotropic states, the measured step-size
tends to the {\sl standard} step-size as $p_1$, $p_2$ become large.
This is precisely what is modelled by ``lattice
refinement''\footnote{Here and throughout we use ``lattice
  refinement'' to refer to any quantisation in which the lattice {\it
    defined with respect to $\mu$} is dynamically varying. In
  particular, then, ``new'' quantisation is simply a special case of
  such ``lattice refinement''.}, albeit in an heuristic way,
 except that we might expect the
``lattice refinement'' to approach the special case of the ``new''
quantisation refinement rate, for large scales.

\section{The three-dimensional case}\label{sec:3D}

In this section, we demonstrate a concrete example of the above
procedure for the case of an anisotropic Bianchi~I model, recently
consistently derived in Ref.~\cite{Ashtekar:2009vc}. In this case the
states are defined on the orthonormal basis
$|\lambda_1,\lambda_2,\lambda_3\rangle$, where $\lambda_i$ are
essentially the scale factors along the three directions of the
Bianchi model.

In the standard approach, isotropic states are taken to be those in
which the dependence on $\lambda_1$,$\lambda_2$ and $\lambda_3$
is ``integrated out'', or more precisely, defining
the volume of the state as $\nu = 2 \lambda_1\lambda_2\lambda_3$ to
eliminate one of the directions ($\lambda_3$), the map
\be
 |\Psi \left( \lambda_1, \lambda_2, \nu\right)\rangle \rightarrow
\Big|\sum_{\lambda_1,\lambda_2} \Psi \left( \lambda_1, \lambda_2,
\nu\right)\rangle \equiv |\Psi \left(\nu\right)\rangle~, 
\ee
produces isotropic states. This projection produces exactly isotropic
system with the equations given in terms of the volume (called the
``$\nu$ quantisation'').  Here we will show that it is also possible
to have a super-selection of states that are isotropic and which
evolve according to an alternative quantisation procedure which may
be modelled by lattice refinement. We are going to work with the three
scale factors $\lambda_i$, to demonstrate that there is an ambiguity
in exactly what the measured volume of such isotropic states would
be. Firstly, we want to find a symmetric superposition of states that
is isotropic, in the sense defined in
Section~\ref{subsec:alternative}.

Consider the state
\be
\label{eq:state1} 
|\tilde\Psi \left( \lambda_1,\lambda_2,\lambda_3\right)
\rangle \equiv \frac{1}{A}\Bigl[ |\Psi \left(
  \lambda_1,\lambda_2,\lambda_3\right)\rangle + |\Psi \left(
  \lambda_3,\lambda_1,\lambda_2\right)\rangle + |\Psi \left(
  \lambda_2,\lambda_3,\lambda_1\right)\rangle \Bigr]~, 
\ee 
with the additional restriction
\beq 
&&\ \ \  \langle \Psi\left(\lambda_1,\lambda_2,\lambda_3\right) 
| \Psi
\left(\lambda_3,\lambda_1,\lambda_2\right)\rangle \nonumber\\&& = \langle
\Psi\left(\lambda_1,\lambda_2,\lambda_3\right) | \Psi
\left(\lambda_2,\lambda_3,\lambda_1\right)\rangle \nonumber \\ && =
\langle \Psi\left(\lambda_3,\lambda_1,\lambda_2\right) | \Psi
\left(\lambda_2,\lambda_3,\lambda_1\right)\rangle\nonumber\\ && \equiv f~,  
\eeq
on the anisotropic states being used.  This additional requirement is
not present in the two-dimensional case, since there is only one
cross-correlator between two anisotropic states, which is trivially
equal to itself. In the three-dimensional system however, discussed here, 
we have three cross-correlators, which need to be equal if the state is 
to be isotropic. This is a restriction that would be present in any
dimension higher than $2$.

The expectation values of the scale factors along each direction of
such a state are
\beq
 \langle \hat{\lambda}_i \rangle = \frac{\lambda_1 + \lambda_2 +
   \lambda_3}{3}~.
\eeq
Note that we have chosen $A$ to normalise the state. It is worth
mentioning that, by choosing a different state, the conditions required
for the expectation values to match change. In particular, for the
state
\beq\label{eq:state2}
|\chi\left(\lambda_1,\lambda_2,\lambda_3\right)\rangle &\equiv&
\frac{1}{\tilde{A}} \Bigl[
 \langle \Psi\left(\lambda_3,\lambda_1,\lambda_2\right)|
 \Psi\left(\lambda_2,\lambda_3,\lambda_1\right)\rangle  
|\Psi\left( \lambda_1,\lambda_2,\lambda_3\right)\rangle
\nonumber \\
&&{\hskip0.05 truecm}
 +\langle \Psi\left(\lambda_1,\lambda_2,\lambda_3\right)|
 \Psi\left(\lambda_3,\lambda_1,\lambda_2\right)\rangle  
|\Psi\left( \lambda_2,\lambda_3,\lambda_1\right)\rangle
\nonumber \\
&&{\hskip0.05 truecm}
+ \langle \Psi\left(\lambda_1,\lambda_2,\lambda_3\right)|
 \Psi\left(\lambda_2,\lambda_3,\lambda_1\right)\rangle  
|\Psi\left( \lambda_3,\lambda_1,\lambda_2\right)\rangle
\Bigr]~,
\eeq
to be isotropic, we require
\beq
&&\ \ \ |\langle \Psi\left(\lambda_1,\lambda_2,\lambda_3\right) 
| \Psi
\left(\lambda_3,\lambda_1,\lambda_2\right)\rangle|^2\nonumber\\
&& = |\langle
\Psi\left(\lambda_1,\lambda_2,\lambda_3\right) | \Psi
\left(\lambda_2,\lambda_3,\lambda_1\right)\rangle|^2 \nonumber \\ && =
| \langle \Psi\left(\lambda_3,\lambda_1,\lambda_2\right) | \Psi
\left(\lambda_2,\lambda_3,\lambda_1\right)\rangle |^2~.
\eeq
With such a state, one also finds that $\langle \hat{\lambda}_i\rangle
=\left( \lambda_1+\lambda_2+\lambda_3\right)/3$, which explicitly
demonstrates the non-uniqueness of the choice of isotropic embedding.

For both these states, the measured scale factor is equal in each
direction and is given by the average of the scale factors of the
underlying, anisotropic states. However, the measured volume of such a
state, found using the volume operator,
\be
 \hat{\nu} |\tilde\Psi\left(\lambda_1,\lambda_2,\lambda_3\right) \rangle =
 2\lambda_1\lambda_2\lambda_3
 |\tilde\Psi\left(\lambda_1,\lambda_2,\lambda_3\right) \rangle~,
\ee
is just $\nu = 2\lambda_1\lambda_2\lambda_3$, which is {\it not} the
cube of the measured scale factor. Thus, whilst it is the eigenvalue
of the anisotropically defined volume operator, it is not necessarily
what we would measure as the volume. Essentially, this is because
whilst the scale factors $\lambda_i$ are measured to be equal in each
direction, they are not eigenvalues of the state, {\it i.e.},
$\hat{\lambda}_i
|\tilde\Psi\left(\lambda_1,\lambda_2,\lambda_3\right)\rangle \neq
\lambda_i|\tilde\Psi\left(\lambda_1,\lambda_2,\lambda_3\right)\rangle$. However,
both the average and the product ({\sl i.e.}, the volume) of the scale
factors are eigenvalues. It is this ambiguity, that leads to the
possibility of deviations from the standard isotropic case. In
particular, for our choice of isotropic states, we have essentially
defined the projection,
\be
\label{eq:projection2}
|\tilde\Psi\left(\lambda_1,\lambda_2,\lambda_3\right)\rangle \rightarrow
\tilde\Psi \Bigl( \langle \lambda\rangle, \nu\left( \langle \lambda \rangle
\right) \Bigr)~, 
\ee 
where the state depends only on the average, $\langle\lambda\rangle$,
of the three scale factors, and their product ({\it i.e.}, the volume) can be
viewed as a complicated function of this average.

The Hamiltonian found in Ref.~\cite{Ashtekar:2009vc} is linear and
can be written as a symmetric combination of three operators that 
act on each $\lambda_1$, $\lambda_2$ and $\lambda_3$ equally. Thus the
Hamiltonian operator satisfies our definition of a ``symmetric operator'',
Eq.~(\ref{eq:sym_op}) and 
hence we can directly apply the above states to it. Alternatively one
can directly evaluate the Hamiltonian acting on a symmetric state
and show show that it contains only states that are themselves symmetric.

 As it is explained in Ref.~\cite{Ashtekar:2009vc}, the Hamiltonian
 acts on uniform steps along the $\nu$ direction, just as in the ``new
 quantisation'' of isotropic cosmologies. Just as in the standard
 case, one finds that the Hamiltonian, acting on symmetric states,
 produces a {\sl super-selection} of points along the $\nu$ direction,
 that are equally spaced, however from Eq.~(\ref{eq:projection2}) it
 is clear that this no longer implies a uniform spacing in either
 $\langle \lambda \rangle$ (the measured scale factor) or $\langle
 \lambda \rangle ^3$ (the measured volume). To deduce the step-size in
 the measured scale factor (which would give the step-size in the
 holonomies that should be used in the case of isotropic LQC), one
 would have to invert the function $\nu \left( \langle \lambda \rangle
 \right)$. The use of ``lattice refinement'' in isotropic LQC is a
 heuristic first approximation to this, and whilst the power law like
 refinements typically used in the literature~\cite{Nelson:2007um} are
 rather crude, they have the advantage of being analytically
 tractable.

There are many ways that such isotropic states can be produced from
superpositions of anisotropic states. Certainly, those given in
Eqs.~(\ref{eq:state1}) and (\ref{eq:state2}) are not the most general
ones, however they serve to demonstrate the possibility that whilst
the underlying, anisotropic system requires specific, well defined
holonomy step-sizes, the isotropic version contains an ambiguity in
the precise value of the step-size that one can take. Here, we have
focused only on the Bianchi~I model, for which a full understanding of
the quantisation exists, however one expects similar effects to occur
when starting from other anisotropic and inhomogeneous models. In
particular, the motivation for lattice refinement in LQC comes from the
fact that one expects that the additional degrees of freedom present
in the underlying, full theory, will not exactly produce the symmetry
reduced models of cosmology. Here we have shown that the set of states
that are measured to be isotropic form a super-selection of states
within the Bianchi 1 model, which have a non-constant measured volume
step. In this sense we have shown that the anisotropic degrees of
freedom of the Bianchi 1 model, that are missing in the symmetric FLRW
case, can produce a non-standard evolution of the FLRW volume, which is
exactly what is modelled by lattice refinement.  It is important to
note however that the Hamiltonian governing the evolution of these
isotropic states is not the standard FLRW Hamiltonian that is usually
considered in lattice refinement models. Thus whilst it may be hoped
that lattice refinement models capture some of the consequences of a
varying discreteness size, the use of the standard Hamiltonian makes
this a rather crude approach.

Finally, it is important to note that in Eq.~(\ref{eq:projection2}),
we chose to view the measured scale factor of the isotropic state as
the primary variable, however one could have chosen to view the
eigenvalue of the volume operator as fundamental, in which case the
isotropic volume of the state would be $\nu$ and the isotropic
Hamiltonian would indeed produce a {\sl super-selection} with uniform
steps in volume. In this case however the scale factor of the state
($\langle \lambda \rangle$) would be related to the volume through a
complicated function (essentially the inverse of $\nu\left(\langle
\lambda\rangle \right)$). This would seem to be largely a matter of
taste, however it is possible that future work may choose one view
point over the other.

\section{Conclusions}

Historically, the quantisation of FLRW cosmologies within LQC,
considered the triad element to be the fundamental variable that is to
be quantised~\cite{Ashtekar:2003hd}.  However, it was realised that
this leads to an instability within the theory at large
scales~\cite{Bojowald:2007ra}, which is particularly problematic for
inflation~\cite{Nelson:2007wj}.  It was later shown that these
problems are cured~\cite{Nelson:2007wj} for the ``new'' quantisation
approach in which the physical volume is considered as the fundamental
object to be quantised~\footnote{Although the onset of inflation
  within LQC may remain problematic, even with the use of the ``new''
  quantisation~\cite{Germani:2007rt}; a question which has been
  recently addressed in Ref.~\cite{aa-ds}.}~\cite{Ashtekar:2006wn}.

In terms of the basic triad component $p$, the discretisation of the
Hamiltonian occurs with step-size given by $p^{{\cal A}}$, where
${\cal A}=0$ for the original quantisation approach and ${\cal
  A}=-1/2$ for the ``new'' quantisation. These two possibilities are
the two limits in which, either only the labels of the edge spins, or
the number of vertices dynamically change in the underlying
lattice (respectively for the ``old'' and ``new'' quantisation
approaches).  Whilst the behaviour of the full Hamiltonian constraint
on a general lattice has yet to be derived from LQG, it is expected
that the constraint on ${\cal A}$ should be $0<{\cal A}<-1/2$, since
the Hamiltonian acts by a combination of both processes. A full
derivation of this from an underlying LQG point of view, is not yet
possible, and here we do not restrict the value of ${\cal A}$ and
refer to the general case as ``lattice refinement''~\footnote{The term
  ``lattice refinement'' is also sometimes used to include only the
  cases with $ -1/2<{\cal A}<0$ case, which explicitly excluded the
  ``new'' quantisation approach. Here, as previously, we use the term
  ``lattice refinement'' in its most general sense, to refer all
  quantisations in which the step-size is given by $p^{\cal A}$, which
  includes both the ``new'' and ``old'' quantisations as special
  cases.}.  There has been much work (and significant debate) on
whether ``lattice refinement'' for a general ${\cal A}$ is
theoretically
consistent~\cite{Ashtekar:2006wn,Corichi:2008zb,Nelson:2008vz} and
strictly working within the symmetry reduced FLRW models, it appears
that only for ${\cal A}=-1/2$ (``new'' quantisation) is this the
case~\cite{Corichi:2008zb}.

However, here we have shown that, given a (fully consistent) quantum
anisotropic model, it is possible to find isotropic states for which
the discrete step-size of the isotropically embedded Hamiltonian
constraint is not (necessarily) that of ``new'' quantisation. This
demonstrates the possibility of ``lattice refinement'' being due to
the degrees of freedom that are absent in the isotropic model. It is
of course possible to embed\footnote{Here the term ``embedding'' is used
in a loose sense. The FRLW model is in fact a ``projection'' of a system
with more degrees of freedom.} the ``new'' quantisation approach within
the anisotropic models and it may be that this is the more natural
procedure, however without an {\it a priori} knowledge of which embedding
process to follow, the conservative approach would be to consider
both.

We have shown that the difference between the two procedures is
essentially due to what one considers to be more fundamental, the
volume of the underlying states ($\nu$) or the measured volume of the
symmetric state ($\langle \lambda \rangle ^3$), which are not
necessarily equal. Choosing $\nu$ leads to the ``new'' quantised
Hamiltonian of isotropic cosmology, whilst choosing $\langle \lambda
\rangle ^3$ results in some kind of different ``lattice
refinement''. This ``lattice refinement'' is significantly more
complicated that the single power law behaviour, $p^{{\cal A}}$,
usually considered, however one may hope to capture some of its
effects using the simple model.

It is important to note that here we have started from a Bianchi~I
model and looked for isotropic embeddings, however similar ideas
should apply more general models. We are not suggesting that the
Bianchi~I model is the underlying geometry of our universe, however it
is sufficient to demonstrate the ambiguities associated with finding
an effective isotropic cosmology within a model that has more degrees
of freedom. This can be the thought of as the source of the ``lattice
refinement'' ambiguity of LQC, however more work is needed to find out
if the simple power law model is a sufficiently accurate approximation
of this ambiguity.  More work is also necessary to see how this new
formulation of ``lattice refinement'', in particular the new
relationship between the effective scale factor, $\langle \lambda
\rangle$, and the volume, $\nu$, effects the consistency of the
isotropic theory.

Finally, one may be concerned with the fact that simply deciding which
variables are the more fundamental one, can have important
implications for the large scale physics of the theory, however this
is not an unusual occurrence for quantum systems. In all quantum
systems, we start from a classical action and follow specific
procedures to produce the quantum version of the theory. Precisely
which classically equivalent action to choose is a source of quantum
ambiguities. From a theoretical point of view, one may hope to have
the choice picked out from a more complete understanding of the
underlying theory, whilst on a practical level, it is often more
useful to parametrise our ignorance and fix these parameters from
experiments and observations. In the latter case one simply assumes
that the approach is an effective theory, approximating some, as yet
unknown, full underlying theory, which is sufficiently accurate for
the current data. In the case of LQC (and indeed classical cosmology)
we are constantly working within the framework of an effective theory
of universal dynamics, in which (it appears) the symmetry reduction is
typically a valid approximation. In this context, ``lattice
refinement'' is simply an effective parametrisation of the fact that
cosmological symmetries are not exact symmetries of the physical
universe. Whilst measuring any consequences of this effective
parametrisation seems unlikely, it is possible that the largest scales
of the Cosmic Microwave Background (which correspond to the earliest
scales to leave the horizon) might retain a signature of this LQC
regime, from which ambiguities such as ``lattice refinement'' may be
fixed.

\vskip.05truecm 
\section*{Acknowledgments}
The work of M.S. is partially supported by the European Union through the
Marie Curie Research and Training Network \emph{UniverseNet}
(MRTN-CT-2006-035863).

\end{document}